# Low-temperature magnetoresistance hysteresis in vanadium-doped $Bi_{2-x}Te_{2.4}Se_{0.6}$ bulk topological insulators

Birkan Düzel[1], Christian Riha[1], Karl Graser[1], Olivio Chiatti[1], and Saskia F. Fischer[1,2,*]

[1] Novel Materials Group, Humboldt-Universität zu Berlin, 10099 Berlin, Germany

[2] Center for the Science of Materials Berlin, Humboldt-Universität zu Berlin, 12489 Berlin, Germany

* Corresponding author

$Bi_{2-x}Te_{2.4}Se_{0.6}$ single crystals show gapless topological surface states and doping ($x$) with Vanadium allows to shift the chemical potential in the bulk band gap. Accordingly, the resistivity, carrier density, and mobility are constant below 10 K and the magnetoresistance shows weak antilocalization as expected for low-temperature transport properties dominated by gapless surface states of so-called three-dimensional topological "insulators". However, the magnetoresistance also shows a hysteresis depending on the sweep rate and the magnetic field direction. Here, we provide evidence that such magnetoresistance hysteresis is enhanced if both three-dimensional bulk states and quasi-two-dimensional topological states contribute to the transport ($x$ = 0 and 0.03), and it is mostly suppressed if the topological states govern transport ($x$ = 0.015). The results are discussed in terms of spin-dependent scattering between the different available states.

# I. Introduction

Topological surface states (TSS) have been convincingly detected by angle-resolved photo emission experiments (ARPES) [1, 2, 3] in materials denoted as topological "insulators" (TIs). Purely topologically-protected transport may be used in novel spintronic or quantum information applications. However, transport measurements often show phenomena based on the co-existence of TSS and bulk states and their contributions are not easily disentangled [4,5]. An example is the low-temperature magnetoresistance hysteresis observed in bulk vanadium-doped $Bi_2Te_{2.4}Se_{0.6}$ single crystals [6]. The causes remain unclear and may originate in the interplay between transport in quasi-two-dimensional topologically-protected surface states and in three-dimensional bulk states.

Here, we investigate ternary TIs $Bi_2Te_{3-y}Se_y$ alloys with a quintuple layer unit of Te–Bi–Se–Bi–Te [7], where the strong bonding suppresses Se vacancies and the Te–Bi anti-site defects [8], which otherwise pose difficulties in the binary TIs, $Bi_2Se_3$ and $Bi_2Te_3$ [9, 10]. $Bi_2Te_{2.4}Se_{0.6}$ has a wide bandgap of 0.25 eV [11] and bandstructure characterization by ARPES has shown gapless TSS for $Bi_{2-x}Te_{2.4}Se_{0.6}$ with addition of vanadium in concentrations $x$ of 1.5 and 3 percent [6]. Due to a change in the chemical potential by doping and an activation energy of about 25 meV, the temperature-dependence of the Van-der-Pauw resistivity, the Hall charge carrier density, and the mobility were shown to be strongly dependent on the vanadium concentration [6]. In general, in high-quality single crystals the low-temperature transport at around liquid-nitrogen temperature is governed by electrons with densities as low as $1.5 \times 10^{16}$ cm$^{-3}$ and mobilities as high as 570 cm$^2$/Vs. At temperatures between 0.3 K and 10 K, the resistivity, carrier density, and mobility remain constant and, conclusively, the transport is dominated by the gapless TSS. This was confirmed by weak anti-localization in the magnetoresistance in the same temperature range. The analysis by the Hikami–Larkin–Nagaoka model yielded phase-coherence lengths of up to 250 nm for the vanadium doping, which fixes the chemical potential in the bulk band gap ($Bi_{2-x}Te_{2.4}Se_{0.6}$ with $x = 0.015$) [6].

However, magnetoresistance measurements in vanadium-doped $Bi_{2-x}Te_{2.4}Se_{0.6}$ single crystals with $x = 0.015$ and $x = 0.03$ revealed a considerable hysteresis depending on the sweep rate and the magnetic field direction [6]. The origins of the hysteresis remain unresolved. This hysteresis occurs at low-temperatures, where the phase coherence length exceeds the elastic scattering length, so that a relation to the quantum correction of the conductivity due to spin-orbit coupling appears to be relevant. It was further shown that the magnetoresistance at a fixed magnetic field increased or decreased exponentially to a static magnetoresistance value with long time constants of several minutes. From this time-dependence a ferromagnetic-like origin of the hysteresis was excluded.

In this study, we aim to discriminate between different sources for such magnetoresistance hysteretic behavior, and investigate $Bi_{2-x}Te_{2.4}Se_{0.6}$ single crystals with three different vanadium doping concentrations of a) $x = 0$, b) $x = 0.015$ and c) $x = 0.03$ by temperature-dependent magnetoresistance measurements.

## II. Experimental Details

The single crystal growth by the Bridgman method, structural and bandstructure characterization have been described previously in detail [7, 12, 6]. The vanadium-doped $Bi_{2-x}Te_{2.4}Se_{0.6}$ single crystals, with $x = 0$, 0.015 and 0.03 were cleaved and cut by a diamond saw into rectangular 2 mm by 5 mm bulk samples with thickness $t_0 = (270 \pm 20)$ µm, $t_{0.015} = (170 \pm 20)$ µm and $t_{0.03} = (650 \pm 60)$ µm. The samples were glued onto a Si substrate with a 1 µm $SiO_2$ layer. Ohmic contacts were formed by wedge bonding Al wires onto the substrate and connecting these to the crystal by Ag paint. All samples were electrically characterized by measuring current-voltage (*I-V*) characteristics. Temperature and magnetic-field dependent measurements were performed in Oxford Instruments *Triton* dilution- and *HelioxVL* ³He-refrigerators, and a CryoVac *KONTI-IT* flow-cryostat. Measurements were performed with a Keithley 2450 Source-Meter Unit or a Keithley 6221 current source with a 2182A nanovoltmeter for *I-V* characteristics, and a Stanford Research System SR830 and Signal Recovery 7265 lock-in amplifiers for low-noise resistance measurements. The resistivity was determined by Van-der-Pauw measurements [13] to $\rho_{VdP} = \frac{\pi \cdot t}{\ln(2)} \cdot \frac{R_1 + R_2}{2} \cdot f$, by measuring resistances $R_1$ and $R_2$ in two different contact configurations and using the thickness *t* and the correction factors *f* of about 0.83. At very low temperatures we obtained the resistivity from the longitudinal resistance $R_{xx}$ in Hall contact configuration, where $\rho_{xx} = R_{xx} \cdot \frac{t \cdot w}{l}$ and *w* denotes the width and *l* is the distance between voltage contacts.

For magnetoresistance measurements a Keithley 6221 current source, a 433 Hz sine signal with a current of 10 nA$_{rms}$ were used. Voltages were measured using the Stanford Research SR560 preamplifier with differential inputs and an SR830 lock-in amplifier. For the crystals with $x = 0.015$ and $x = 0.03$ measurements of $R_{xx}$ were possible to obtain the magnetoconductance. For the crystal with $x = 0$ measurements had to be performed in Van-der-Pauw geometry due to the loss of one contact.

# III. Results

## A. Temperature-dependent resistance measurements

The temperature-dependent resistivity, determined from *I-V* characteristics with a maximum current of 5 mA at zero magnetic field in Van-der-Pauw geometry, are shown in Figure 1a). All *I–V* curves show ohmic characteristics, as depicted in the inset for the example of the $x = 0.015$ sample at 10 K. However, the two different configurations in Van-der-Pauw geometry show differences which cannot be explained by purely geometrical means for an isotropic homogeneous current distribution. It resembles an inhomogeneous current distribution in the bulk crystal which is typically expected for TIs with TSS. This is more so the case if bulk bands contribute to the transport, because the chemical potential is in or near the bulk conduction band edge. From the resistance measurement during cooling and warming of the $x = 0.015$ crystal, there are no significant effects observed from thermal cycling. The samples all appear stable with respect to different cooling cycles and ageing.

For temperatures above 40 K all samples show a semiconducting-like temperature dependence of the resistivity $\rho$ determined from Van-der-Pauw measurements, see Fig. 1a). However, for temperatures below 20 K the crystal with $x = 0.015$ shows a saturation resistance for lower temperature. Such is also reached below 10 K for the cases of no doping ($x = 0$) or a higher doping concentration ($x = 0.03$), as can be seen from Fig. 1b).

Measuring the differential resistivity by lock-in amplifiers at very small excitation currents of only 10 nA, we may exclude heating effects when approaching 50 mK base temperature. In the ultra-low temperature range from 2 K to 50 mK we observe a nearly constant resistivity, which strongly indicates the dominant influence of the TSSs in the transport behavior for all doping cases.

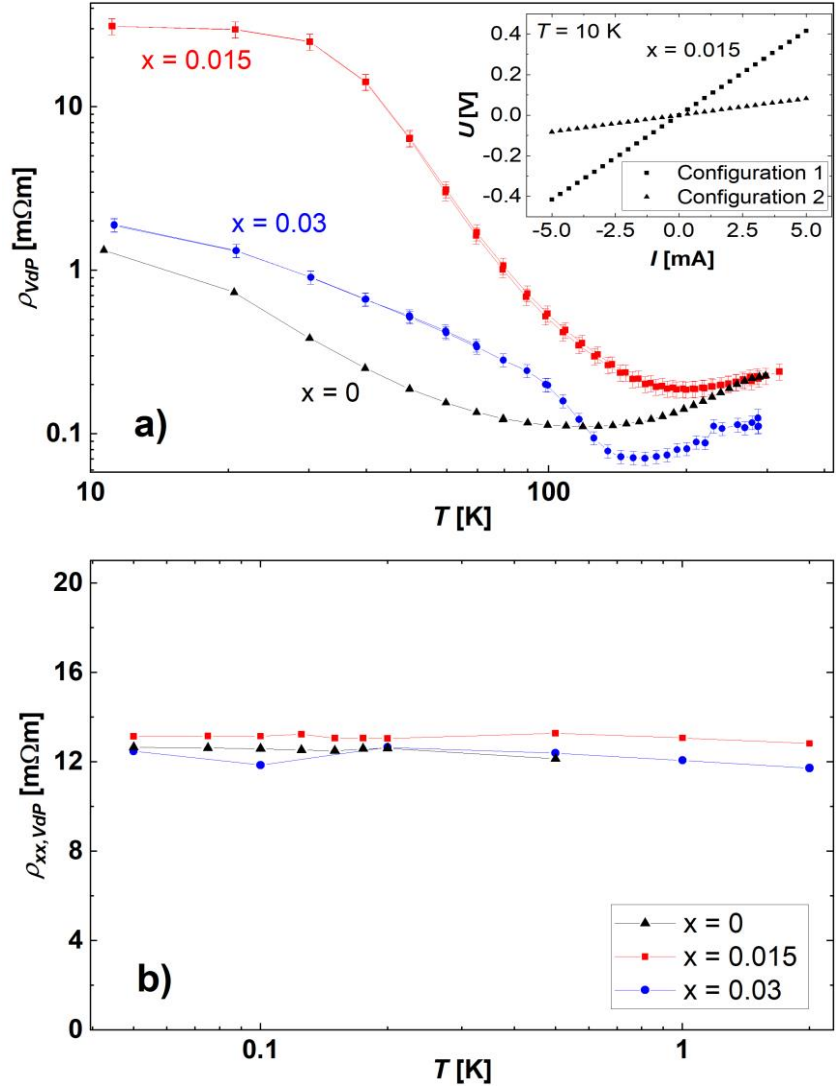

**Figure 1.** Temperature-dependent resistivity of $V_x Bi_{2-x} Te_{2.4} Se_{0.6}$ bulk crystals with $x = 0$, $x = 0.015$ and $x = 0.03$. a) The resistivity was obtained from Van-der-Pauw measurements; the inset depicts the current-voltage characteristics in the crystal with $x = 0.015$ at 10 K for two Van-der-Pauw configurations. b) Nearly constant resistivity below 2 K signals the dominance of transport in the topological surface states at ultra-low temperatures for all crystals, independent of the vanadium concentration. Longitudinal resistance configuration was used for $x = 0.015$ and $x = 0.03$ and Van-der-Pauw configuration for $x = 0$.

### B. Weak antilocalization

Magnetoresistance $\rho(B)$ was measured in the temperature range from 50 mK to 2 K, with a magnetic field $B$ perpendicular to the crystals. The magnetoconductivity $\Delta\sigma = (1/\rho(B)) - (1/\rho(0))$ is given in Fig. 2 and shows for all three samples the magnetic-field dependence expected from weak antilocalization [14]: This quantum correction results in a maximum in the conductivity $\sigma$ for $B = 0$, which decreases with increasing field and with increasing temperature. The maximum occurs due to spin-orbit coupling, which affects the phase-coherence in the diffusive transport. The crystal with $x = 0$ shows a change of conductivity of $|\Delta\sigma| \approx 14\, e^2/h$ over a magnetic field range of 50 mT, which decreases for $T > 150$ mK. The crystal with $x = 0.015$ shows a similar change with $|\Delta\sigma| \approx 12\, e^2/h$ over 50 mT, which decreases for $T > 125$ mK. However, the crystal with $x = 0.03$ shows a change of $|\Delta\sigma| \approx 8\, e^2/h$, which does not change significantly up to $T \approx 1$ K. This indicates that the vanadium-doping changes the rate of scattering mechanisms that affect the phase-coherence of the charge carriers.

The magnetoconductivity curves in Fig. 2 can all be fitted using the Hikami-Larkin-Nagaoka (HLN) equation [15] given by

$$\Delta\sigma = \sigma(B) - \sigma(0) = \alpha \frac{e^2}{2\pi^2\hbar}\left[\ln\left(\frac{B_\varphi}{B}\right) - \Psi\left(\frac{1}{2} + \frac{B_\varphi}{B}\right)\right], \tag{1}$$

where $\alpha$ is a prefactor and $B_\varphi = \hbar/4el_\varphi$ a characteristic magnetic field. The phase-coherence length $l_\varphi$ obtained from the HLN fits is depicted in Fig. 3.

All three samples show a saturation of $l_\varphi$ for $T \to 0$, which decreases monotonically with increasing vanadium concentration: $l_{\varphi,0} \approx 300$ nm, 260 nm and 170 nm for $x = 0$, 0.015 and 0.03, respectively. For $x = 0$ and 0.015 $l_\varphi(T)$ decreases for temperatures above approximately 200 mK, while for $x = 0.03$ it is nearly constant up to 1 K. The phase coherence lengths at temperatures above 200 mK agree with data obtained previously [6]. The vanadium doping reduces monotonically $l_\varphi$ and the change $|\Delta\sigma|$ over a given field range, which implies that it increases inelastic scattering or introduces additional magnetic scattering [14].

The prefactor $\alpha$ is expected to be –1/2 for WAL in a single 2D electron system [15], however we obtain from the HLN fits values that far exceed this value. Similar results were obtained in the WAL analysis of $Bi_2Se_3$ microflakes [4], where it was found that bulk states contributed significantly to the transport as a stack of 2D layers. The values of $l_\varphi$ are not affected by $\alpha$.

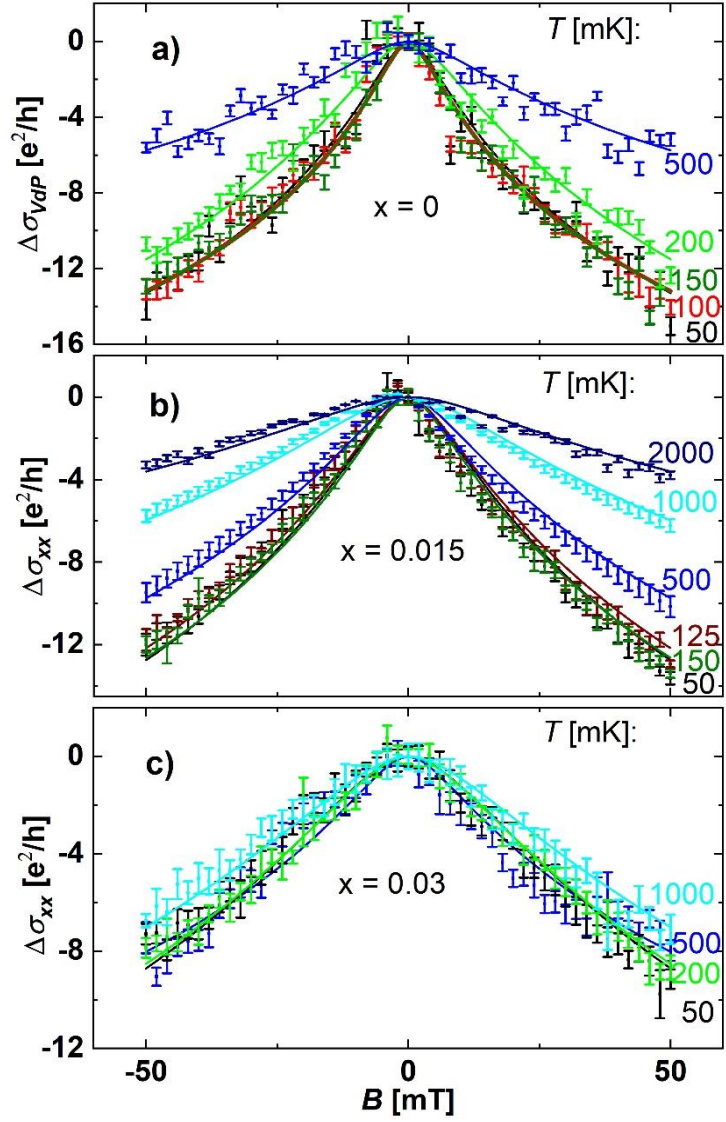

**Figure 2.** Magnetoconductivity of $V_xBi_{2-x}Te_{2.4}Se_{0.6}$ bulk crystals with a) $x = 0$, b) $x = 0.015$ and c) $x = 0.03$ at temperatures between 50 mK and 2 K, as depicted in the figure. Solid lines represent the fitted HLN equation. The value of $\Delta\sigma_{xx}$ in b) and c) was obtained from the longitudinal resistance configuration and the rectangular-cut geometry of the crystals. The value of $\Delta\sigma_{VdP}$ was obtained from the Van-der-Pauw configuration. The slope in a) and b) decreases with increasing temperature. The temperature dependence in c) is less pronounced.

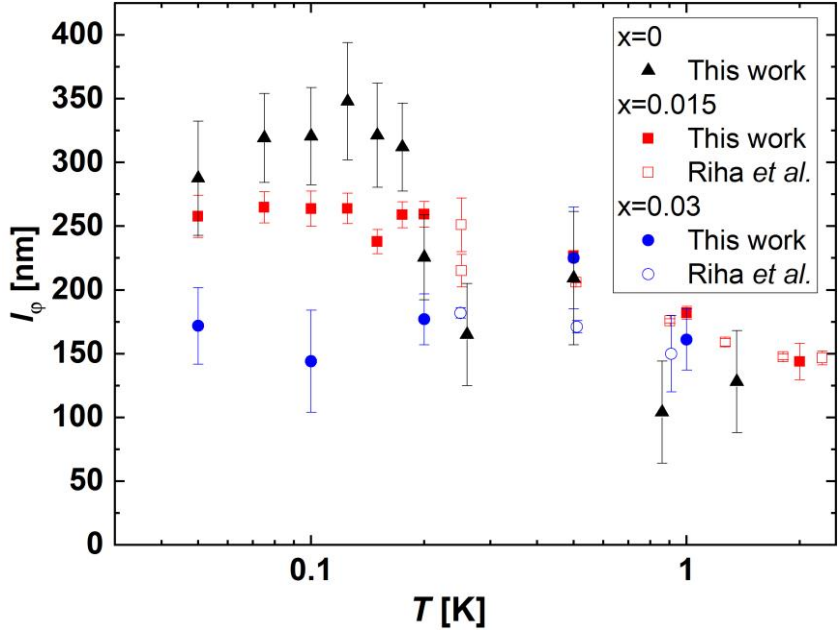

**Figure 3.** Phase-coherence $l_\varphi$ of $V_x Bi_{2-x} Te_{2.4} Se_{0.6}$ with $x = 0$ (black triangles), $x = 0.015$ (red squares) and $x = 0.03$ (blue circles), extracted from the HLN fits at different temperatures. All samples show a nearly constant $l_\varphi$ below 200 mK. Filled symbols have been obtained from measurements in the Triton dry dilution refrigerator. The results agree with measurements in the HelioxVL ³He-refrigerator (open symbols) [6] for temperatures between 200 mK and 2 K. A reduction of the constant phase-coherence length with increasing vanadium concentration is observed below 200 mK.

### C. Sweep-rate dependent magnetoresistance hysteresis

In accordance with the *sweep-rate* dependent hysteresis in the magnetoresistance in higher magnetic fields in vanadium-doped $Bi_{2-x} Te_{2.4} Se_{0.6}$ single crystals with $x = 0.015$ and $x = 0.03$ [6], we confirm that it occurs for all crystals, as shown in Fig. 4. However, there are two additional observations: First, a magnetoresistance hysteresis occurs also for the $x = 0$ crystal for temperatures up to 0.8 K, which directly validates that vanadium does not play a decisive role in the hysteresis. Second, the observed hysteresis is much more prominent and appears equivalent for $x = 0$ and $x = 0.03$ (Fig. 4a and c), than for $x = 0.015$ (Fig. 4b). This indicates that the position of the chemical potential with respect to the bulk band gap is relevant: Less pronounced hysteretic behavior is observed where the chemical potential is situated in the band gap and TSS provide the dominant states for transport processes ($x = 0.015$, Fig. 4b). A substantially more pronounced hysteretic behavior in the magnetoresistance is observed when the chemical potential is situated in or near the conduction band (edge), so that TSS *and* bulk states contribute substantially to the transport ($x = 0.03$, Fig. 4c). Both cases were unambiguously distinguished by ARPES measurements [6].

As can be seen from Fig. 4, the magnetoresistance hysteresis is more pronounced for an increased sweep rate. Whether the magnetic field is increased or decreased in a sweep determines if the magnetoresistance is suppressed or raised, respectively, with respect to the static magnetoresistance. Returning to zero-magnetic field the magnetoresistance values coincide independently of the magnetic field direction and the decrease to the static value at zero-field requires about 500 s. This long time constant points to relaxation phenomena by scattering mechanisms from a non-equilibrium (excited) state to thermal equilibrium.

In general, for sweep-rate dependent magneto-hysteresis effects several causes may be discussed as possible origins. Here, we exclude a general Joule heating relevant for high-current densities, because our excitation current of 10 nA is low with respect to the ultra-low temperatures due to the macroscopic size of the samples. This also holds if we consider only transport in TSS confined to surface regions, because the surface areas of all crystals are macroscopic. Similarly, we may exclude any hysteretic magnetoresistance effects due to long-range spin ordering as may occur in ferromagnets, antiferromagnets or otherwise spin-ordered materials or in the vicinity of such.

Previously, a spin-dependent scattering for eddy currents in the TSSs was suggested [6]. While this may play a role, the present results indicate that it does not suffice as the origin of the sweep-rate dependent hysteresis, as observed in Fig. 4. Here, the strongly enhanced hysteresis occurs when the chemical potential allows occupation of TSS *and* bulk states (Fig. 4a ($x = 0$) and c ($x = 0.03$)) and the hysteresis is much smaller when the chemical potential is in the band gap (Fig. 4b ($x = 0.015$)). However, the notion that the higher surface conductivity in the TSSs gives rise to enhanced eddy currents becomes important to establish a spatially inhomogeneous non-equilibrium distribution in the occupation of states by the time-varying magnetic field. In particular, the topological nature of the TSS due to the spin-orbit coupling may lead to a topological magneto-electric effect [18]. This itself will not be observable in the present samples, however, it may give rise to spatial redistribution of charges (polarization by induction) and a shift in the quasi-chemical potential of the different TSS involved (c. f. Fig. 5, 10 and 11 and Sec. II.B.2 in Ref. [10]).

In the case where both TSS and bulk states are available for scattering processes to restore equilibrium, the mixture of causes for different spin-dependent scattering rates results in different time scales for up- and down-sweeps of the magnetic field. The contributions may arise from spin-momentum-locking for electrons in TSS, different degree of spin polarization by the magnetic field in the TSS and bulk states, and different number of Landau levels in the TSS and the bulk. Regarding the latter case, the number of Landau levels is very large in both TSS and bulk states, but differ by a factor of thousand based on the electron densities.

Evidence for differences in the relaxation times between the cases of contributing TSS and bulk states versus mainly TSS may be the mobilities as estimated from the overall field dependence in Fig. 4. The magnetoresistance for $x = 0$ and $x = 0.03$ shows a $B^2$-dependence for $|B| > 0.75$ T (see Fig. 4a and c). A fit with the semi-classical magnetoresistance $\Delta\rho/\rho(0) = (\mu B)^2$ for diffusive transport [16] yields a drift mobility of $\mu \approx 2100$ cm$^2$/Vs and $\mu \approx 1600$ cm$^2$/Vs for $x = 0$ and $x = 0.03$, respectively. This drift mobility is much larger than the Hall mobility obtained from a one-band analysis from low-field magnetotransport. The magnetoresistance for $x = 0.015$ does not show a clear $B^2$-dependence in the same magnetic field range (see Fig. 2b), but a linear dependence. This signals a competition between the WAL cusp and the semi-classical $B^2$-dependence for diffusive transport [17], which is dominated by the TSS in this magnetic field range.

While the details of the mechanisms leading to the hysteretic behavior require further investigation, the present data are indicative that the occurrence of magnetoresistance hysteresis in similar 3D TI materials without long-range spin order point to a substantial contribution of bulk states to scattering processes which restore a non-equilibrium distribution to equilibrium. Its appearance or the lack thereof may be taken as a signature of the position of the chemical potential with respect to the bulk band edge in cases where gating or ARPES measurements are not feasible.

Future experiments on the thickness dependence by employing thin epitaxial films or single crystalline flakes may help to clarify the role of bulk states in restoring non-equilibrium situations in the presence of topologically-protected surface states.

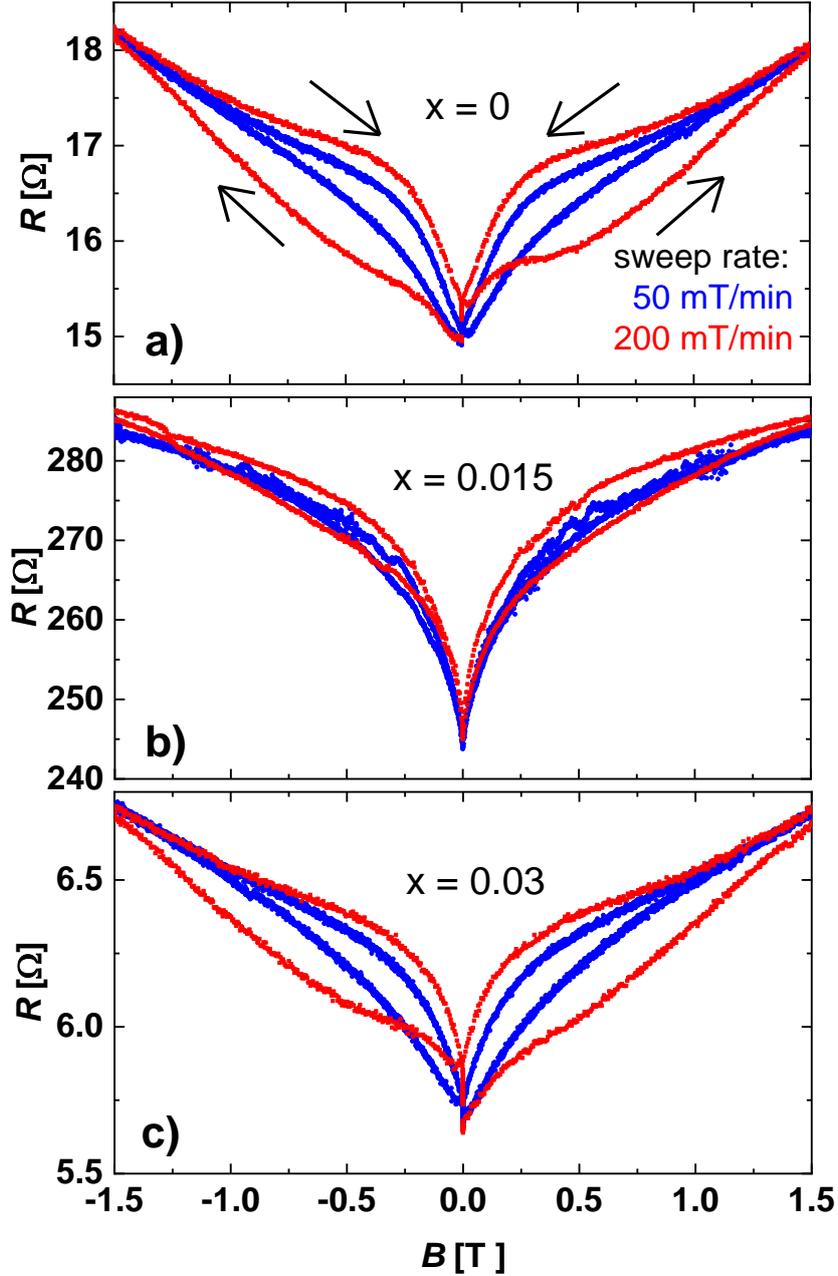

**Figure 4.** Magnetoresistance for two different sweep rates of the magnetic field (50 mT/min (blue curves) and (200 mT/min (red curves)) of $V_xBi_{2-x}Te_{2.4}Se_{0.6}$ bulk crystals with a) $x = 0$, b) $x = 0.015$ and c) $x = 0.03$ at 250 mK. The magnetic field is varied as indicated by the arrows in a), leading to a hysteresis which increases for increasing sweep rate. The hysteresis is most prominent for crystals in which the chemical potential lies in the bulk band, see a) and c). It is strongly suppressed where the chemical potential is in the band gap and topological surface states dominate the transport, see b).

# IV. Conclusion

Analysis of transport measurements of bulk $Bi_{2-x}Te_{2.4}Se_{0.6}$ single crystals with vanadium doping ($x$) shows the dominance of topologically-protected surface states by the temperature-independent resistivity below 10 K and by weak antilocalization in the magnetoconductivity. The appearance of a sweep-rate dependent hysteresis in the magnetoresistance in fields up to 1.5 T is most pronounced when the position of the chemical potential is in or near the bulk conduction band edge, so that both three-dimensional bulk states and quasi-two-dimensional topological states contribute substantially to scattering processes. The origin of the hysteresis can be explained by the difference in spin-dependent scattering rates in surface states and in the bulk. Non-equilibrium may be evoked by induction due to time-varying fields in the topological surface states, so that scattering is required to restore the equilibrium distribution in static magnetic fields. Application of magnetoresistance hysteresis investigations can open a route to identify the optimal doping in topological insulator materials.

**Acknowledgements**

The authors thank Prof. Oleg E. Tereshchenko, Novosibirsk State University, for the supply of the material and thank Dr. Evangelos Golias, Dr. Jaime Sánchez-Barriga and Prof. Oliver Rader, Helmholtz-Zentrum-Berlin für Materialien und Energie, for providing information on the band profile by ARPES measurements, and Jürgen Sölle, Humboldt-Universität zu Berlin, for technical support in the sample preparation. The authors gratefully acknowledge the partial financial support from the Deutsche Forschungsgemeinschaft under grant INST 276/709-1 and by the priority programme "Topological Insulators: Materials – Fundamental Properties – Devices" SPP 1666, and the JAMA Lab.

## Author Declarations

**Conflict of Interest**

The authors have no conflicts to disclose.

**Author Contributions**

**Birkan Düzel**: Conceptualization (supporting); Data curation (equal); Formal analysis (equal); Investigation (equal); Methodology (equal); Validation (equal); Visualization (equal); Writing – Original Draft Preparation (equal); Writing – Review & Editing (equal). **Christian Riha**: Data curation (equal); Formal analysis (equal); Investigation (equal); Validation (equal). **Karl Graser**: Formal analysis (equal); Investigation (equal); Validation (equal). **Olivio Chiatti**: Conceptualization (supporting); Data curation (supporting); Formal analysis (supporting); Methodology (supporting); Visualization (supporting); Writing – Original Draft Preparation (equal); Writing – Review & Editing (equal). **Saskia F. Fischer**: Conceptualization (equal); Funding acquisition (lead); Methodology (equal); Project Administration (equal); Supervision (lead); Visualization (equal); Writing – Original Draft Preparation (equal); Writing – Review & Editing (equal).

## Data Availability Statement

The data that support the findings of this study are available from the corresponding author upon reasonable request.